\documentstyle[12pt,aasms4]{article}
\slugcomment{To appear in ApJ Letters}
\begin{document}
\title{ Detection of the Li I 6104 \AA~  transition
in the  Population II star HD 140283\footnote{Based on observations
collected at the European Southern Observatory, La Silla, Chile}}
\author{ Piercarlo Bonifacio} 
\affil{Osservatorio Astronomico di Trieste, Via G.B. Tiepolo 11, 34131, Trieste}
\authoremail{bonifaci@oat.ts.astro.it}
\author{Paolo Molaro}
\affil{Osservatorio Astronomico di Trieste, Via G.B. Tiepolo 11, 34131, Trieste}
\authoremail{molaro@oat.ts.astro.it}

\begin{abstract}
{
Lithium is one of the few primordially produced elements.
The value of the primordial Li is 
taken to be that observed in metal--poor dwarfs, 
where  it is  not  contaminated by
stellar Li sources which act on longer time scales.
The atmospheric abundance is currently derived  from the  
LiI $\lambda\lambda   
6707 \AA~$  resonance transition and the validity
of the models employed has been questioned \markcite{k95} (Kurucz 1995).
In this letter we report the first detection of the Li I
  $\lambda\lambda   6104 \AA~ 2^2P - 3^2D$  subordinate transition in  
the prototype  population II star HD~140283.
 The same Li abundance of (Li/H) $=1.4\times 10^{-10}$  
is found consistent with both the 
 resonance and subordinate lines. 
The two lines    form at different depths
in the atmosphere implying  that the 1-D homogeneous atmospheric models
 used in the abundance determination are 
essentially correct.
When
coupled with the standard big bang yields, the  Li in the halo dwarfs 
provides two solutions for 
the baryon-to-photon ratio
$\eta_{10}= n_{b}/n_{\gamma} \times 10^{10}$ and for
the present baryon density $\Omega_b h_{70}^2=0.0748\eta_{10}$: 
a) a first solution  at  $\eta_{10}\approx 1.8$, 
consistent with   the $\eta_{10}$ implied by the
high deuterium values 
$D/H\approx 2\times 10^{-4}$ 
observed in some quasar absorption systems
\markcite{webb} (Webb et al 1997) and b) a second 
solution  at $\eta_{10}$ $\approx$  4 which is  consistent, within the errors,
with    the low  deuterium D/H =$3.4\times 10^{-5}$
measured in other quasar  
absorption systems\markcite{burles} (Burles \& Tytler 1998).
}
\end{abstract}

\keywords{cosmology: observations --- line: formation --- 
nucleosynthesis --- stars: abundances ---
stars: individual (HD140283) ---
stars: Population II}

\section{Introduction}

Lithium,  together with D and $\rm ^{3,4}He$, 
 is one of the few elements produced by nuclear reactions
in the first minutes after big bang\markcite{wfh}
(Wagoner, Fowler \& Hoyle 1967). The observations of these
elements and their extrapolation to the primordial values are 
consistent with the predictions of 
the standard primordial nucleosynthesis providing,
together with the relic radiation and the expansion of the Universe,
a robust support to the big bang theory.
Recently, additional support to the primordial nature of Li in halo dwarfs 
has come
from the observations of Li in 
metal-poor stars of the
thick disc\markcite{mbp97} (Molaro, Bonifacio \& Pasquini 1997).
This population is
chemically and kinematically distinct from the halo, but 
has  the same Li abundance of the halo.
Minniti et al (1997) 
\markcite{min97} claimed detection of Li, at the plateau level,
in
a metal-rich, but old star, belonging
to the Galactic Bulge. 
Finally Li at the plateau level has also been detected in a star
which was possibly born in an external
galaxy and then accreted   by the Milky Way\markcite{mol97}
(Molaro 1997).

So far the Li abundance 
has been  always obtained only from the analysis of the Li I
 $\lambda\lambda$ 6707 \AA~ 
resonance doublet. This is not a very comfortable situation 
in the light of the importance of the  determination of
lithium  abundances in stars 
for primordial nucleosynthesis,  stellar structure and chemical evolution.

Our ability to determine
the Li abundance using simple plane-parallel homogeneous atmospheres,
 has been recently debated\markcite{k95,kis97,gp97} (Kurucz 1995;
Kiselman 1997; Gadun \& Pavlenko 1997).
The analysis of several  lines, which sample different
depths in the stellar atmosphere is crucial to test the correctness
of the modelling. The one dimensional, homogeneous, static models which are
currently employed may arise concern because they ignore
the fine structure and hydrodynamic phenomena such as granulation
which are seen on the Sun.

The Li I $\lambda\lambda$ 6707 \AA ~resonance transition is the only 
one readily
available to spectroscopic observation.
The strongest subordinate line at 6104 \AA ~ is  much  fainter 
and blended with Fe I line and has been so far detected only in young 
T Tauri stars (Hartigan et al 1989)\markcite{hart} and 
Li-rich giants (Merchant 1967, Wallerstein \& Sneden 1982)
\markcite{merchant,ws}, 
were  Li is more than about 1 dex more abundant 
owing to the Galactic Li production.
\par
In this letter we report the detection of the Li I $\lambda\lambda$ 
6104 \AA ~transition in the spectrum of the metal--poor star HD140283.
Both this line and the resonance line are consistent with the computations
made using a one dimensional, homogeneous model atmosphere, thus
increasing our confidence that 
this model represents a satisfactory average of the complex
fine structure expected in metal--poor stars.

The use of Li observed in halo dwarfs as an indicator of primordial abundance
rests on the absence of any Li  depletion. 
Depletion is predicted
by non-standard models which take into account rotational mixing
\markcite{pin92} (Pinsonneault, Deliyannis \& Demarque 1992)
 or diffusion
\markcite{vau95} (Vauclair \& Charbonnel 1995),  
but these models predict a downturn
of the hot side of the Li plateau and considerable dispersion. 
It seems that neither the downturn nor the large dispersion
is present in the observations, which
suggests that diffusion or rotational mixing do not affect
significantly 
the Li observed at the stellar surface of metal--poor dwarfs.
However,
the downturn can be very small ($\approx 0.2$ dex)
for the purely diffusive case and a suitable
choice of the mixing length parameter ($\alpha=1.5$, see
Deliyannis et al 1990\markcite{deli90}) and 
the issue of intrinsic dispersion remains rather controversial
with
some positive claims. 
Ryan et al (1996)\markcite{ryan} identify a triplet of stars 
(G064-012, G064-037, CD -33$^\circ$ 1173)
with similar
colors, but different Li abundances by a factor of 2.5.
Then there is  the case of star BD+23 3912 
which has a [Fe/H]$\approx -1.3$ to  $-1.5$
and  a Li abundance which is about 0.20-0.36 dex higher than 
the plateau (Rebolo et al 1988,
King et al 1996\markcite{reb88,king96}). Moreover
Boesgaard et al (1998)\markcite{boesgaard}
find differences of up to $\approx 0.5$ dex
among seven subgiants of M92 but the same objects show other chemical 
peculiarities, namely [Mg/Fe] is 0.55 dex lower 
and [Na/Fe] is 0.76 dex larger than in HD140283 (King et al 1998)
\markcite{king98} . 

\section{Observations and analysis}

We observed HD140283 on August 10$^{\rm th}$ and
11$^{\rm th}$    1997 at La Silla, Chile, with the
ESO New Technology Telescope and the EMMI spectrograph
under sub-arc-second seeing conditions. The 
high incidence angle echelle grating (tan $\theta = 4$) and
a  projected slit width  of $0.8 ''$,
provided  a 
resolution of $\lambda /\Delta \lambda \approx$ 61000, as measured from
the Th lamp emission lines in the  region around 6104  \AA .
The spectra were reduced in a standard way and then coadded yielding a 
 S/N
of $\approx$ 360.
The coadded spectrum was normalized by fitting a spline
through points determined by averaging the spectrum over
continuum windows identified with spectrum synthesis.
Figure 1  shows  the region around the LiI $2^2P-3^2D$,
6103.6 \AA~ transition. 
\par
The Li I feature is
clearly detected at 6103.6 \AA , red-wards  of the 
 Ca I 6102.723 \AA~  line. The equivalent width of the feature is
1.8$\pm 0.3$ m\AA~ and the detection is  at 6$\sigma$ of
confidence level.

In Fig 1 and 2 the LiI 6104 \AA ~and the LiI 6707 \AA ~doublet
are shown superimposed on synthetic spectra. 
The synthetic spectra were computed using the SYNTHE code \markcite{k93} 
(Kurucz 1993)
assuming a  Li abundance  
(Li/H)=1.4$\times 10^{-10}$ \markcite{bm}
(Bonifacio \& Molaro 1997),  not derived from 
present data.  The model atmosphere for HD 140283 is
the same used in Bonifacio \& Molaro (1997)
\markcite{bm}
and has parameters  
$\rm T_{eff}= 5691$, log g =3.35 , [Fe/H]=-2.5,
microturbulent velocity of $\rm \xi=1~kms^{-1}$.  
This iron abundance is close to the value derived by
King et al (1998) 
\markcite{king98}
who find
[Fe/H]=-2.58 
with an effective temperature ($\rm T_{eff}=5650$)  close to ours.
The synthetic spectra were broadened  with 
a gaussian instrumental point spread
function of 5 kms$^{-1}$, derived from the Th-Ar lines,
and then  
with a rotational profile  of  4 kms$^{-1}$,
derived from the Li I 6707 \AA ~
resonance line.
The subordinate transition was synthesized as three lines,
$2^2P_{1/2}-3^2D_{3/2}$, $2^2P_{3/2}-3^2D_{5/2}$, 
$2^2P_{3/2}-3^2D_{3/2}$. 
The corresponding wavelengths\markcite{mw76} (Martin \& Wiese 1976),
marked in Fig. 1 as vertical bars, and $\log gf$
values\markcite{ln} ({Lindg\aa rd} \&  {Nielsen} 1977)
 are: 6103.538 \AA, 0.101, 6103.649 \AA, 0.361,
6103.664 \AA, -0.599.
\par

\section{Discussion}

The LiI line forming regions lie  in the upper 
part of the atmospheric convective
zone where Li is mostly ionized due to its low ionization potential.
This is why the determination of precise Li abundances 
requires accurate observations, accurate stellar effective temperature  
and an appropriate
modeling of the atmosphere of a metal poor  star.
The model-atmospheres employed are one dimensional (1-D), with
plane parallel geometry and ignore any inhomogeneity effect,
such as granulation.
Qualitative computations,  based on a two-stream model atmosphere,
suggested that the abundance of Li in halo dwarfs 
could be underestimated by as much as a factor of 10\markcite{k95}
(Kurucz 1995), 
but more 
recent calculations based on 2-D\markcite{gp97}
(Gadun \& Pavlenko 1997) and 3-D\markcite{kis97} (Kiselman 1997)
atmospheric models
show that effects of granulation on the LiI lines are much less important.
Granulation effects in the atmosphere have a depth dependence  and this should 
produce different effects in the resonance and subordinate doublets.

As  can be 
seen from  figures 1 and 2, the same Li abundance reproduces satisfactorily 
both the 6104 \AA ~and the 6707 \AA
~doublets. 
The two transitions form at different depths in the stellar atmosphere:
unit optical depth at wavelength 6707.761 \AA~ is attained at
$log (\tau_{Ross})\approx -0.57$,
corresponding in our model to a local temperature
of 5235 K, while at wavelength 6103.649 \AA~ it is already
attained at $log (\tau_{Ross})\approx -0.09$, or T=5915 K.
The resonance line receives contributions from a more extended region
than the subordinate line. Unit optical depth
at the  wavelength
at which the residual intensity is 0.999, 
is attained at $log (\tau_{Ross})\approx -0.11$ for the resonance line,
but at $log (\tau_{Ross})\approx -0.08$ for the subordinate line.
Thus the subordinate line samples deeper and hotter layers than
the resonance line, as shown in Fig. 3.
\par

The lower level of the Li 6104 \AA~ transition is the upper 
level of the 6707 \AA $2^2S-2^2P$~ transition.
Our synthetic spectra are 
computed under the LTE assumption and
the  consistency between the two lines  
implies a correct computation of  the populations of the 
2S, 2P and 3D 
levels. This is in agreement with the 
theoretical
estimations which predict relatively small corrections for 
NLTE effects in the LI 6707 \AA ~line \markcite{carl94,pavmag}
(Carlsson et al 1994; Pavlenko \& Magazz\`u 1996). 
Thus the  detection of a subordinate LiI line,
and its consistency with the abundance derived from the
resonance 6707 \AA~ doublet, provides 
support to the correctness of this Li abundance.

The consistency of 
the abundances based on the LiI 6707 \AA~ and 6104 \AA~  transitions
observed in HD 140283 supports  the Li abundances measured in 
the population II stars, using 1-D model atmospheres, in the last decades. 
The new generation of large  telescopes will
allow  to measure the 
Li 6104 \AA ~Li I  subordinate doublet in other much fainter
population II stars, thus permitting  to verify  this consistency
on the grounds  of  a statistically  significant  sample,
and ultimately achieve a more accurate measurement of the
primordial Li abundance.
\par
Among the light elements produced in the first minutes 
after the big bang, Li is the only one which
shows a non monotonic behaviour with $\eta_{10}$, 
the so-called  {\sl Li-valley}, 
which reflects the different nuclear 
reactions which synthesize Li at different baryonic
densities. 
The most recent measurement of the Li primordial abundance is 
 $\rm (Li/H)=1.73 \pm 0.05_{stat} \pm 0.2_{sys} 
\times 10^{-10}$\markcite{bm}
(Bonifacio \& Molaro 1997), which is the mean value of 41 
halo stars for which precise 
effective temperatures, 
determined by means of the infrared flux method \markcite{alonso}
({Alonso}, {Arribas}, \& {Martinez-Roger}  1996),
were
available. 
The 
systematic errors, which dominate the error budget,   come 
from a possible offset of $\pm$ 75 K 
in the zero point of the temperature 
scale of cool stars.
This Li abundance intercepts
the primordial yields for two different values of $\eta_{10}$, 
which unfortunately 
do not help in resolving the deuterium 
and helium controversies. 
Each  solution for $\eta_{10}$ obtained
from Li is consistent with either
the high-deuterium/low-helium\markcite{webb,olive}
(Webb et al 1997; Olive, Steigman \& Skillman 1997)
 or the low-deuterium/high-helium
\markcite{burles,izo97} (Burles \& Tytler 1998; Izotov, Thuan
\& Lipovetsky 1997).
The lower $\eta_{10}$ 
requires considerable D destruction
to match the 
presently observed abundance in the local interstellar medium of the Galaxy. 
The higher $\eta_{10}$ value is also consistent with
the low deuterium (D/H $=3.9 (\pm 1)\times 10^{-5}$)
derived 
from the 92 cm 
hyperfine transition emission towards the 
unprocessed Galactic anti-center 
\markcite{chen97} (Chengalur, Braun \& Butler Burton 1997).

\clearpage
\figcaption{The observed Li I $2^2P-3^2D$ transition (solid line) and the
synthetic spectrum (dotted line), computed with the parameters 
$ T_{eff}= 5691$, log g = 3.35 , [Fe/H]=-2.5 and 
(Li/H)=1.4$\times 10^{-10}$.
}
\figcaption{
The observed Li I $2^2S-2^2P$ resonance transition (solid line)
and the synthetic spectrum (dotted line), computed with the 
parameters given in the caption to Fig. 1. 
}
\figcaption{Temperature structure of our model atmosphere for HD 140283,
as function of the Rosseland optical depth. The asterisks indicate
the convective layers. The horizontal line marks the depths at which
the monochromatic optical depth $\tau_\lambda = 1$, across the profile of the
Li I 6707 \AA ~line. The square symbol marks the depth at which
$\tau_{6103.649}=1$, the range over which unit optical depth is attained
across the profile of the subordinate line
 is smaller than the size of the symbol. 
}
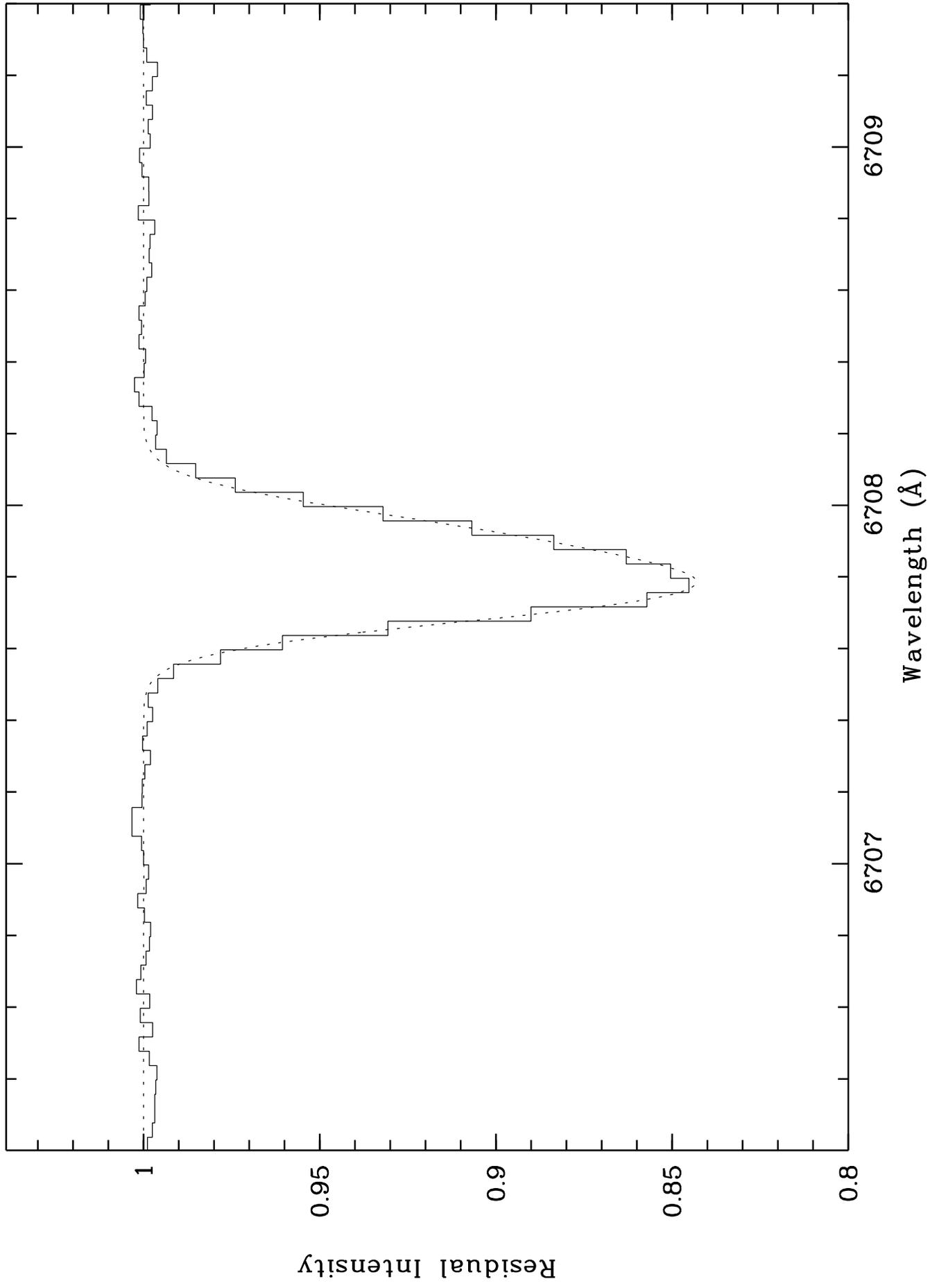
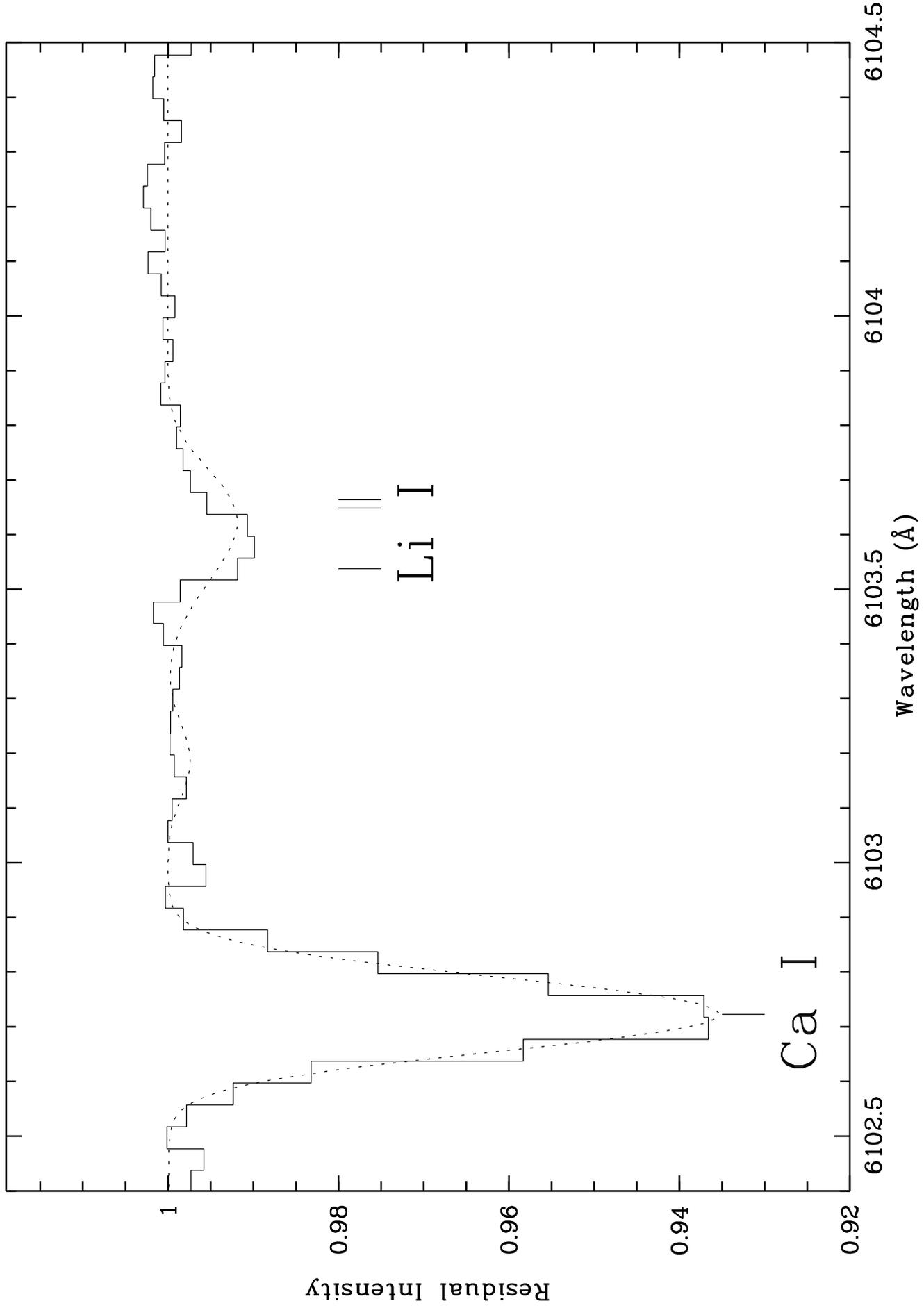
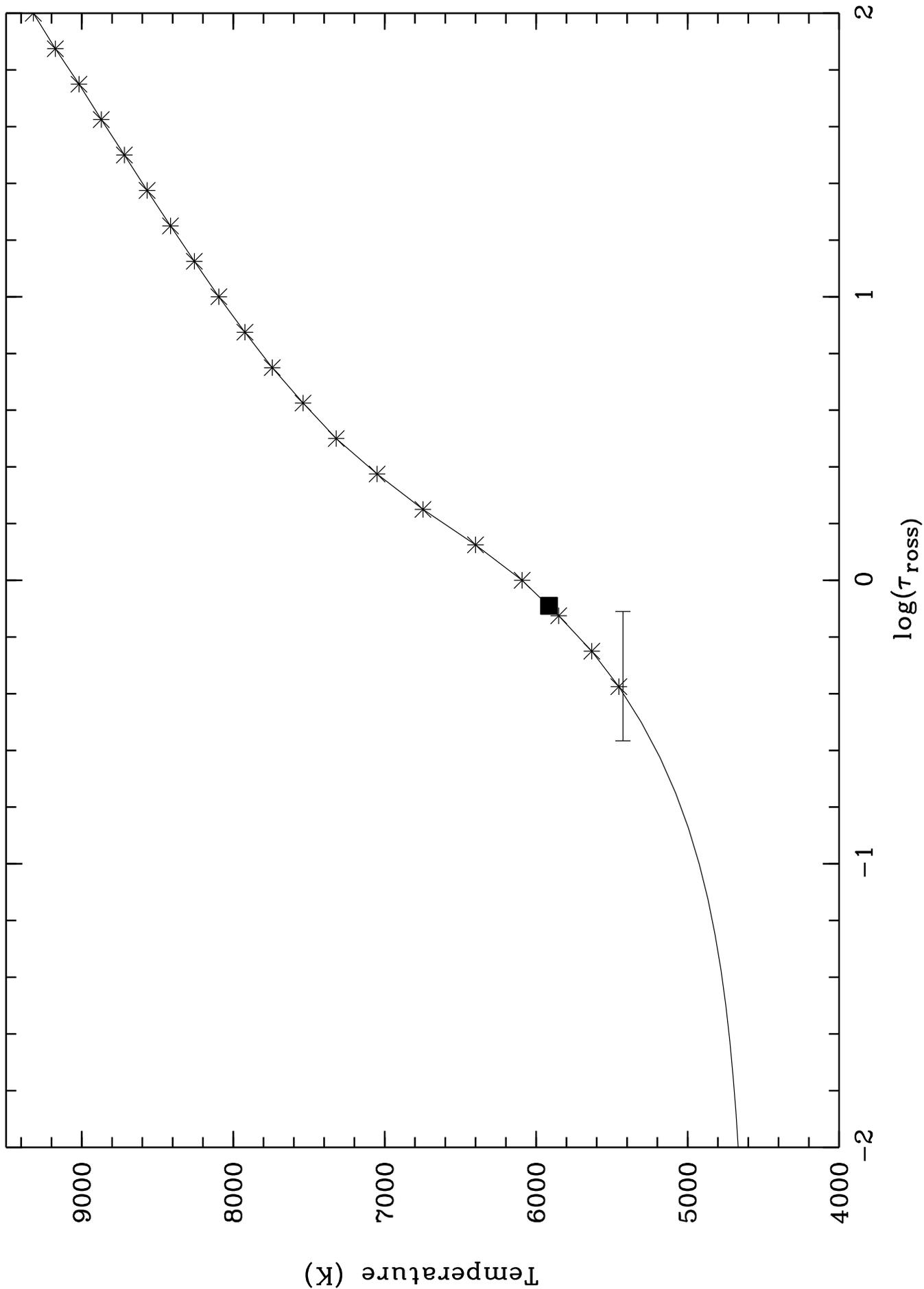
\end{document}